\documentclass[fleqn,onecolumn,10pt]{wlscirep}
\usepackage{graphicx}
\usepackage{dcolumn}
\usepackage{bm}
\usepackage{hyperref}
\usepackage{xcolor}
\usepackage{soul}
\usepackage{tikz}
\usepackage{csquotes}

\usepackage[makeroom]{cancel}
\usepackage{comment}

\title{Unveiling homophily beyond the pool of opportunities}
\author[1,2]{Sina Sajjadi}
\affil[1]{Complexity Science Hub Vienna, Vienna, Austria}
\affil[2]{Central European University, Vienna, Austria}
\affil[3]{Graz University of Technology, Graz, Austria}

\author[1]{Samuel Martin-Gutierrez}

\author[1,3]{Fariba Karimi}

\begin{document}

\begin{abstract}
Unveiling individuals' preferences for connecting with similar others (choice homophily) beyond the structural factors determining the pool of opportunities, is a challenging task. Here, we introduce a robust methodology for quantifying and inferring choice homophily in a variety of social networks. Our approach employs statistical network ensembles to estimate and standardize homophily measurements. We control for group size imbalances and activity disparities by counting the number of possible network configurations with a given number of inter-group links using combinatorics. This method provides a principled measure of connection preferences and their confidence intervals.
Our framework is versatile, suitable for undirected and directed networks, and applicable in scenarios involving multiple groups.
To validate our inference method, we test it on synthetic networks and show that it outperforms traditional metrics. Our approach accurately captures the generative homophily used to build the networks, even when we include additional tie-formation mechanisms, such as preferential attachment and triadic closure. Results show that while triadic closure has some influence on the inference, its impact is small in homophilic networks. On the other hand, preferential attachment does not perturb the results of the inference method.
We apply our method to real-world networks, demonstrating its effectiveness in unveiling underlying gender homophily. Our method aligns with traditional metrics in networks with balanced populations, but we obtain different results when the group sizes or degrees are imbalanced. This finding highlights the importance of considering structural factors when measuring choice homophily in social networks.
\end{abstract}

\maketitle

\section*{Introduction}

Homophily, our tendency to connect with similar others \cite{mcphersonBirdsFeatherHomophily2001}, is one of the most fundamental mechanisms of human interaction. This mechanism has dramatic consequences on societies and individuals, leading to unequal access to financial \cite{separate} and social resources \cite{bischoff_segregation_2019} and impacting the economy of the cities \cite{stier2022effects, stier2023city}. Homophily can perpetuate stereotypes \cite{carter_you_2017}, affecting perception biases \cite{leeHomophilyMinoritygroupSize2019} and health disparities \cite{beck_color_2020, sajjadi2022structural, manna2023social, hiraoka2022herd}.

Measuring homophily has long been a central endeavor in multiple fields, such as sociology \cite{bojanowski2014measuring,lawrence2020homophily}, computer science \cite{lim2021large,veldt2023combinatorial} and network science \cite{newmanMixingPatternsNetworks2003,cantwellMixingPatternsIndividual2019}. Researchers have identified several distinct notions of homophily, ranging from choice homophily to outcome homophily \cite{mcphersonBirdsFeatherHomophily2001,lawrenceHomophilyMeasuresMeaning2020}. Therefore, using an appropriate measurement that aligns with the specific notion one aims to assess is essential to obtain meaningful and interpretable results.

In the context of social networks, homophily can be considered as a disproportionate occurrence of links between nodes belonging to the same social groups. This phenomenon is usually called \emph{outcome} or \emph{observed} homophily. However, the intuitive concept of homophily is usually tied to some notion of personal preference (or \emph{choice homophily} \cite{mcphersonHomophilyVoluntaryOrganizations1987}), outcome homophily arises from a combination of mechanisms, where individual preference is just one factor. Other determining factors are structural constraints \cite{feldFocusedOrganizationSocial1981,marsdenCrosscuttingSocialCircles1985}, as the relative size of groups in the pool of potential contact opportunities can induce outcome homophily even if the links are made at random (this is sometimes called \emph{baseline homophily} \cite{mcphersonBirdsFeatherHomophily2001}). Additionally, other tie formation mechanisms besides choice homophily can indirectly induce homophilous associations, such as the preference to connect with popular people (\emph{preferential attachment} \cite{karimiHomophilyInfluencesRanking2018,avinHomophilyGlassCeiling2015a}), or the inclination to befriend your friends' friends (\emph{triadic closure} \cite{peixotoDisentanglingHomophilyCommunity2022,asikainenCumulativeEffectsTriadic,goodreauBirdsFeatherFriend2009}).

While measuring outcome homophily is straightforward, the multiple confounding elements discussed above make the accurate measurement of choice homophily a real challenge. Choice homophily metrics identify biases towards in-group linking by controlling for sources of homophily unrelated to personal preference. Some are based on the deviation of linkage rates from a baseline model of random association \cite{colemanRelationalAnalysisStudy1958,newmanMixingPatternsNetworks2003, wang2023gender}. Others assume a (log)linear model between the inter-group link proportions and unknown latent variables that depend on the groups and their relationships \cite{marsdenHomogeneityConfidingRelations1988,goodmanSimpleModelsAnalysis1979}. 
However, as we will demonstrate, current homophily inference methods do not effectively consider the pool of structural opportunities when assessing choice homophily. Additionally, these methods do not provide information about the uncertainty or margin of error in homophily measurements.

In this work, we develop statistical models of network ensembles that incorporate parameters for inter-group linkage bias, enabling us to infer choice homophily. Our inference models are based on counting the number of possible network configurations conditioned on a given number of inter-group links weighted by each group's association preferences. They control for the effect of relative group sizes and activity disparities and are robust with respect to the presence of preferential attachment and mostly robust in the presence of triadic closure. Furthermore, they can deal with asymmetric situations where both homophilic and heterophilic tendencies might coexist. By controlling for such effects, the intra-group linkage bias parameters can then be directly identified as choice homophily, facilitating the interpretation of the results. We further derive analytical measures for assessing the uncertainty of the inference.

The proposed approach in this paper is based on statistical inference methodology. Similar approaches to studying various aspects of connection preferences and dynamics have been used in recent years. For example, Peixoto \cite{peixotoDisentanglingHomophilyCommunity2022} applied this framework to disentangle the effects of triadic closure from homophily and community structure, Cantwell et al. \cite{cantwellMixingPatternsIndividual2019} inferred distributions of homophilic preferences for individual nodes, and Altenburger et al. \cite{altenburgerMonophilySocialNetworks2018} utilized this framework to measure the degree of similarity among an individual's friends.

\section*{Results}

\subsection*{Illustrating choice vs outcome homophily} \label{sec:outcom-vs-choice}

Choice homophily is the intrinsic preference of agents to interact with similar others, while outcome homophily represents the observable outcome of such interactions\cite{mcphersonHomophilyVoluntaryOrganizations1987,
asikainenCumulativeEffectsTriadic,
doi:10.1287/orsc.2018.1208}.
The relationship between choice homophily and outcome homophily is not always one-to-one or linear, as outcome homophily arises from the interplay of several mechanisms in addition to choice homophily.

A significant factor influencing the difference between choice and outcome homophily is the relative size of the interacting social groups. For simplicity, let us consider a system with only two social groups, one with more members than the other, such that we have a (numeric) \emph{majority} group and a \emph{minority} group. Even in a random case, all agents naturally experience greater exposure to majority members, leading to higher outcome homophily among the majority and lower outcome homophily among the minority members.

However, the disassociation between choice and outcome homophily in networks cannot be solely attributed to population disparity. Inequality in the number of potential social connections also plays a significant role \cite{fischer_dwell_1982, marsden_core_1987, moore_structural_1990, andersson_higher_2018}.
These disparities are evident in real-life situations, where lower socioeconomic status individuals have fewer neighborhood connections in a network, and social mobility can exacerbate these disparities, with those experiencing downward mobility having even fewer non-kin associations \cite{van_groenou_network_2003,jahani2023long}. These variations can be attributed to inequality in \emph{social resources} \cite{kawachi2000social, burt_network_2000, lin2008network, coleman1988social}. Individuals with higher social status have access to a wider network due to a higher number of weak ties \cite{granovetter_strength_2023}, which can be leveraged to form more friendships and connections.
In addition to socioeconomic status, social skills can impact popularity and the number of friends one has \cite{gottman_social}.
Age, marital status, and parental status are other factors that can impact an individual's number of friends in a network \cite{gillespie_close_2015}.
This disparity in potential connections or degrees within the social network causes some groups to have more connections than others. As a result, one group gains greater visibility among others, leading to outcome homophily values that differ from what would be predicted solely based on the choice homophily.

To illustrate how disparities in group size and social resources affect the discrepancy between outcome and choice homophily, we use a network model that generates synthetic networks with arbitrary values of choice homophily. In this model, the generative choice homophily corresponds to a parameter ${h}^{\mathrm{(gen)}}$ that determines the intrinsic tendency for in-group linking. In order to consider the disparities in social resources, we consider link formation activity as the average number of links established by each node in a group. For illustrative purposes, we study a simple system with two social groups, tunable group size, and link formation activity. We then use the External-Internal (EI) index \cite{EI-measure} as stated in Eq. \ref{eq:EI} to calculate the outcome homophily for these networks, comparing the resulting values with the choice homophily, ${h}^{\mathrm{(gen)}}$, that is a known parameter of the model. The EI index is a straightforward measure for homophily that quantifies how much more frequently links are formed between similar others ($I$, internal) compared to links between different others ($E$, external) in the network.
The EI index ranges from $-1$ (full homophily) to $+1$ (full heterophily):

\begin{equation} \label{eq:EI}
EI = \frac{E-I}{E+I}
\end{equation}

We generate synthetic social networks with two types of nodes (majority and minority) and different choice homophily values using the following network growth model:

\begin{enumerate}
    \item The network begins with two connected nodes of random types.
    \item New nodes join the system. With probability $f_a$, the node is of type $a$ (minority) and otherwise of type $b$ (majority).
    \item The new node establishes $k^r_{init}$ links to older nodes, where $r$ can be either $a$ or $b$, recognizing that groups can exhibit varying activity rates. An older node of the same type will be chosen with a probability proportional to ${h}^{\mathrm{(gen)}}$ and of a different type with a probability proportional to $1-{h}^{\mathrm{(gen)}}$.
\end{enumerate}

\begin{figure*}[!ht]
	\begin{center}
		\includegraphics[width=0.9\linewidth]{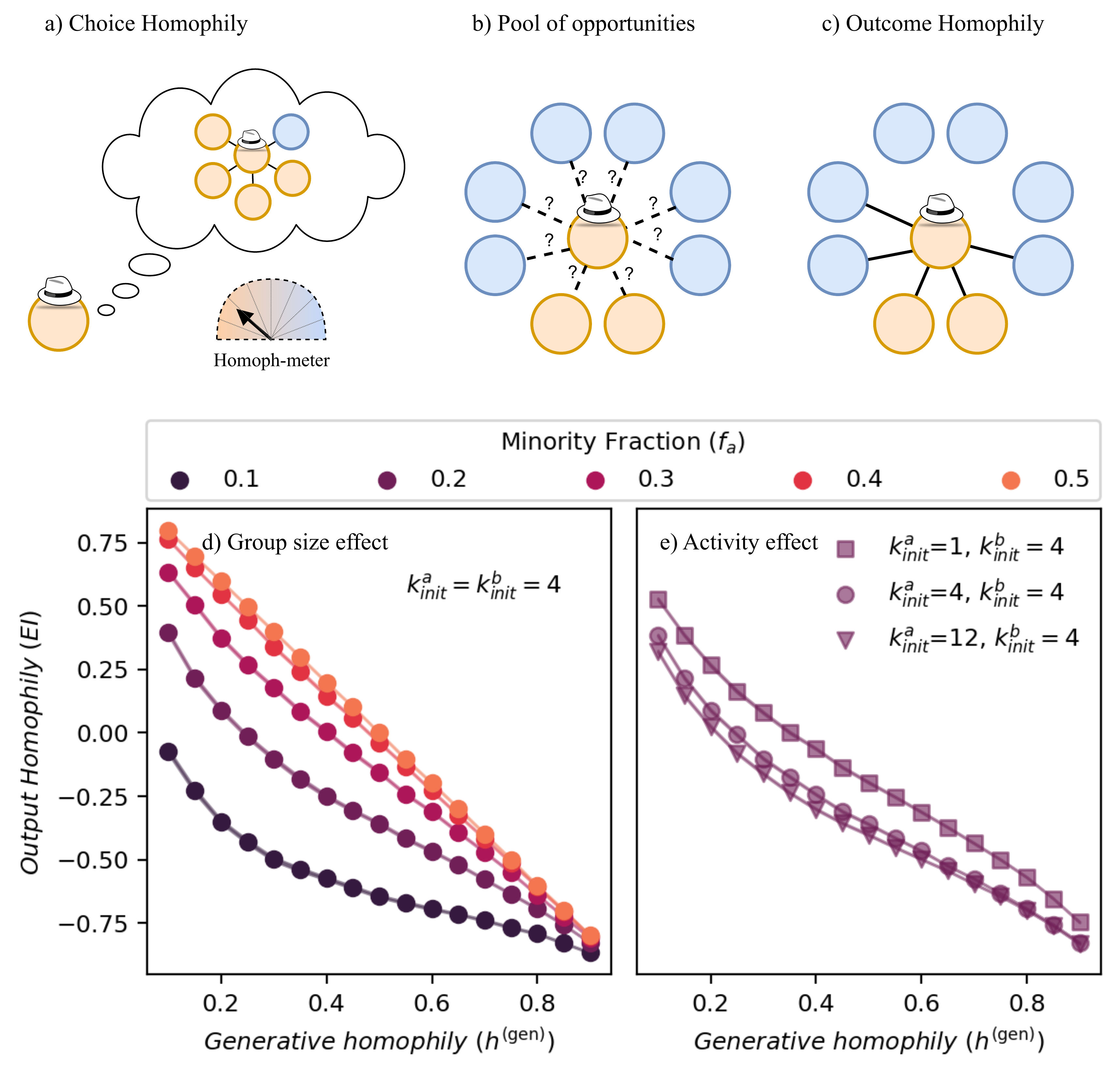}
		\caption{
  \textbf{Relationship between choice and outcome homophily.}
a) The hatted agent is highly homophilic, so they prefer to interact with agents of the same color (orange).
b) However, there is a larger population of blue agents in the area. 
c) Consequently, the hatted agent has more different (blue) friends than similar (orange) ones, resulting in low outcome homophily.
d) Relationship between the generative homophily parameter used to build synthetic networks and the outcome homophily EI measured under different group sizes. Color denotes the fraction of the minority group.
e) Relationship between the network generative homophily and the EI measure under heterogeneous activity rates for a fixed minority fraction of $0.2$. Symbols indicate the number of links created by each new minority node.
In both plots, the x-axis indicates the generative homophily value ${h}^{\mathrm{(gen)}}$, and the y-axis represents the outcome EI measure.}

		\label{fig:EI}
	\end{center}
\end{figure*}

Figure \ref{fig:EI} depicts the disparity between choice homophily, ${h}^{\mathrm{(gen)}}$, and EI homophily index. The top row shows an agent who is more interested in interacting with alters of the same kind, as illustrated by the thought bubble and the homoph-meter preference. However, among the available alters, the agents of the same type are the minority (panel b). Hence, the agent will have fewer same-type connections (panel c) than she would like according to her choice homophily. 

Figure \ref{fig:EI} bottom row shows that the relationship between choice and measured homophily depends both on the minority group fraction (panel d) and the activity/degree disparity (panel e). Specifically, the relationship is only linear (as we would expect) when group sizes are balanced and have the same activity level. Thus, outcome homophily is not solely determined by choice homophily but also by other factors like group size and activity. Hence, we need a measure capable of disentangling these other effects in order to calculate the choice homophily given the network data.

This issue is not limited to the EI measure; in Supplementary Information Section \emph{\ref{sec:coleman-outcom-vs-choice}} we show that Coleman's index, a standard homophily index originally conceived as a choice homophily metric, has similar shortcomings\cite{colemanRelationalAnalysisStudy1958}.

In the remainder of this section, we present our mathematical framework for inferring choice homophily in undirected networks involving two groups. Afterward, we conduct comprehensive tests using synthetic networks to validate our method. Next, we extend our framework to directed networks.
We then introduce a canonical ensemble approximation to improve computational tractability.
Finally, we demonstrate the practical utility of our model by applying it to real-world empirical networks.

\subsection*{Mathematical Framework}

To infer the choice homophily or the intrinsic preference of agents to interact with agents of similar/different types, we devise a robust method based on statistical network ensembles.
This method requires the network structure and group memberships (nodes' attributes) for all nodes as inputs. In our framework, we consider attributes that often remain fixed under the duration of our analysis, such as ethnicity, race, and gender.

As outlined before, the population imbalance and activity disparity (manifested as the node degree) across groups can affect the observed share of in-group and out-group interactions. Our method primarily disentangles these effects to infer the underlying choice homophily of the agents.

Our method is inspired by the configuration model \cite{newman2010networks}, which we use as a null model. In the configuration model, the number of nodes and the degree of each node are kept fixed, and new networks are built by shuffling the available links.
The intuition is that if there is a homophilic (or heterophilic) bias in a given empirical network, the observed networks would have fewer (or more) inter-group links than expected according to the null model.

\begin{figure*}[!ht]
	\begin{center}
		\includegraphics[width=\linewidth]{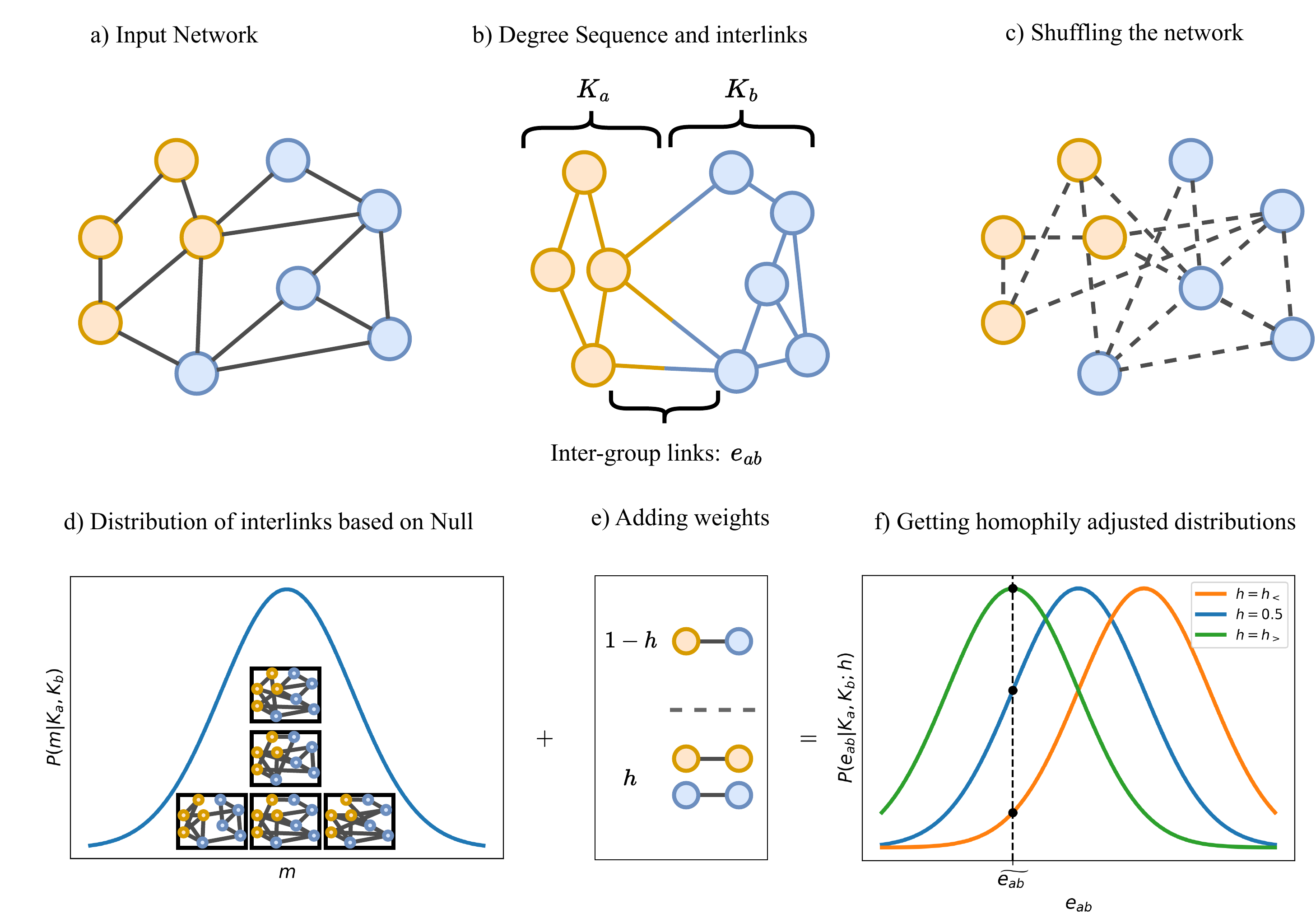}
		\caption{\textbf{Illustration of the mathematical framework.} a) Input network, with nodes of two types.
            b) Counting the number of half-links of each type ($K_a, K_b$) and inter-group links ($\widetilde{e_{ab}}$).
            c) Reshuffling the network based on $K_a, K_b, \widetilde{e_{ab}}$.
            d) Computing the probability distribution of ${e_{ab}}$ given $K_a, K_b$. Each configuration represents a reshuffled network based on the null model. Some ${e_{ab}}$ values have more possible configurations than others.
            e) Adding probability weights $h$ and $1-h$ to homophilic and heterophilic links.
            f) Computing the $h$ adjusted probability distributions. Colors indicate homophily value. The dashed line marks the empirical number of inter-group links $\widetilde{e_{ab}}$. Black dots mark the probability of the empirical value under each distribution. In this scenario, the green distribution is considered the most \textit{likely}.
}
		\label{fig:Illustration}
	\end{center}
\end{figure*}

In the following, we outline the steps of our method for the most basic scenario, where we have an undirected network composed of only two types of nodes, labeled $a$ and $b$ (see Figure \ref{fig:Illustration}.a):

\begin{enumerate}
    \item We define $K_r$ as the total number of half-edges (stubs) originating from nodes of type $r$ where $r$ can be either $a$ or $b$.
    We further define $e_{rs}$ as the number of links going from block $r$ to block $s$ when $r \neq s$ and double the number of links within block $r$ when $r = s$ \cite{peixoto2017nonparametric} (only for undirected networks). For the case of two blocks $a$ and $b$, $e_{ab}$ indicates the number of inter-group links connecting nodes of different types.
    (as depicted in Figure \ref{fig:Illustration}.b).
    \item We fix the values of $K_r$ and rewire the network by pairing the half-stubs with each other (as shown in Figure \ref{fig:Illustration}.c), exhaustively identifying all the networks that can be built with the given $K_r$ and grouping them by the number of inter-group links $e_{ab}$. However, instead of performing this computation with simulations or any algorithm of network manipulation, we use combinatorics to count the number of possible configurations that result in a given ${e_{ab}}$. This includes considering the number of link arrangements within each group and the number of ways to pair up the $e_{ab}$ inter-group links, given the total number of stubs $K_r$ for each community $r$.
\end{enumerate}

Assigning equal probability to each possible network configuration and neglecting multi-edges and self-loops \cite{newman2010networks}, we can calculate $P(e_{ab}|K_a, K_b)$, the probability distribution for the number of inter-group links ${e_{ab}}$, using Eq. \ref{eq:neutral-prob-dist} (illustrated in Figure \ref{fig:Illustration}.d). 

\begin{equation} \label{eq:neutral-prob-dist}
P(e_{ab}|K_a, K_b) = \frac{ \Omega(e_{ab}|K_a, K_b) }{ \Omega(K_a, K_b) } = \frac{ e_{ab}! \binom{K_a}{e_{ab}} \binom{K_b}{e_{ab}} (K_a - e_{ab}-1)!! (K_b-e_{ab}-1)!!  }{ (K_a + K_b -1) !! }
\end{equation}

Where $\Omega(e_{ab}|K_a, K_b)$ denotes the number of configurations resulting in $e_{ab}$ inter-group links, given the total number of stubs of the communities, $K_a$ and $K_b$, and $\Omega(K_a, K_b)$ denotes the total number of configurations, given $K_a$ and $K_b$ (with any number of inter-group links).
As the total number of links of each group is exactly conserved in all realizations of this ensemble, we call it a microcanonical model. A canonical alternative will be introduced in Section \hyperref[sec:canonical]{\emph{Canonical Ensemble Equivalent}}.
The mathematical derivations for the microcanonical ensemble can be found in the Supplementary Information Section \emph{\ref{SI:mathematics}}.

Since the connections are random and unbiased, we have a theoretical probability distribution of $e_{ab}$, the number of inter-group links theoretically expected \textit{given no group preference}.
One way to quantify choice homophily in a real network would be to compare the empirically observed value of $e_{ab}$, which we define as $\widetilde{e_{ab}}$, with its expected value given $K_a$ and $K_b$ through a p-value, determining whether the empirical network exhibits homophily (low left-tail p-value), heterophily (low right-tail p-value) or neutrality (high p-value). For example, a similar approach has been employed in \cite{wang2023gender}.

However, we go one step further by parameterizing homophily in the model. For this purpose, we assign different probability weights $h$ and $1-h$ to the intra and inter-group links (Figure \ref{fig:Illustration}.e).
Hence, a particular network configuration with $E = \frac{K_a + K_b}{2}$ total links, ${e_{ab}}$ inter-group links and $E-e_{ab}$ intra-group links will be assigned a probability $\frac{(1-h)^{e_{ab}}  h^{(E - e_{ab})}}{\mathcal{C}}$, where $\mathcal{C}$ is the normalization constant.
If $h > 0.5$ ($h < 0.5$), the homophilic (heterophilic) links would be over-represented. A value of $h=0.5$ recovers the neutral case already discussed.

Using this scheme, we obtain a new probability distribution $P(e_{ab}|K_a, K_b; h)$ also dependent on $h$ (Figure \ref{fig:Illustration}.f), which controls the homophily of the network in the configuration-like generative network model: 

\begin{equation}\label{eq:prob-h-adjusted}
P({e_{ab}}|K_a, K_b; h) = \frac{ \Omega({e_{ab}}|K_a, K_b; h) }{ \Omega(K_a, K_b; h) } = \frac{ {e_{ab}}! \binom{K_a}{{e_{ab}}} \binom{K_b}{{e_{ab}}} (K_a - {e_{ab}}-1)!! (K_b-{e_{ab}}-1)!! (1-h) ^ {e_{ab}} h ^ {(E - {e_{ab}})} }
{ \sum_{e_{ab}} \Omega({e_{ab}}|K_a, K_b; h) \delta_{mod(K_a - {e_{ab}},2), 0} }
\end{equation}

Where $\delta_{x,y}$ indicates the Kronecker delta with values $1$ when $x=y$ and $0$ otherwise.
We use the Kronecker delta $\delta_{mod(K_a - {e_{ab}},2), 0}$ because only inter-group link counts with the same parity as $K_a$ are acceptable. Otherwise, there would be an odd number of intra-group stubs, and pairing would not be possible. $K_a$ and $K_b$ are then guaranteed to have the same parity, as the total $K_a + K_b$ is even in any possible network.

Inferring homophily in an empirical network is thus equivalent to determining the most plausible value of $h$. To this end, we reinterpret the probability distribution in Eq. \ref{eq:prob-h-adjusted} as a likelihood measure $L(h|K_a, K_b; \widetilde{{e_{ab}}})$ of a network having homophily $h$ given the observation of $\widetilde{{e_{ab}}}$ inter-group links.
The maximization of this likelihood function provides us with the most likely value of homophily $h$ of the network.
We illustrate this process in Figure \ref{fig:Illustration}.f, where we show three possible distributions with their respective homophily values. Clearly, the green curve has a higher likelihood of describing the empirical value of $\widetilde{{e_{ab}}}$. Hence, we assign the network with the homophily value corresponding to the green curve.
Given the probability distribution of the inferred homophily, this methodology will directly provide us with a confidence interval for the inferred value.
While this section solely focused on the case with two groups, the generalization to a higher number of groups is straightforward; see Supplementary Information Section \emph{\ref{section:SI-more-groups}}. We have also generalized the model to directed networks with potentially asymmetric homophilic tendencies, which we will discuss in section
\hyperref[sec:directed_general]{\emph{Asymmetric Homophily Preferences}}.

\subsection*{Testing the mathematical framework with synthetic networks}

To assess the performance of our method, we use synthetic networks with tunable values of generative/input choice homophily ${h}^{\mathrm{(gen)}}$. In the following section, we feed such networks to our inference framework and recover the value of $h$. The matching between these values would indicate the success of our inference method.

\subsubsection*{Varying Group Fractions and Activity Rates}

\begin{figure}[!ht]
	\begin{center}
		\includegraphics[width=0.8\linewidth]{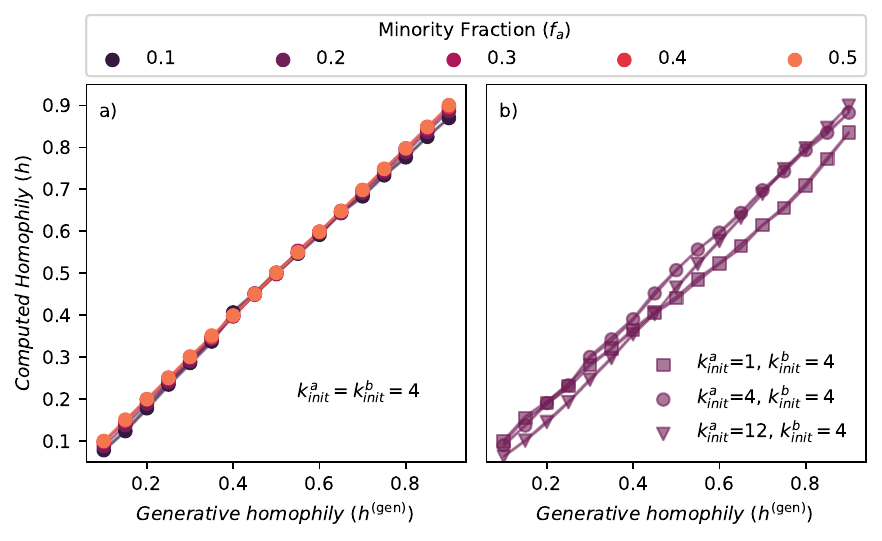}
		\caption{\textbf{Evaluating the performance of our inference method using the growth model with homophily.} x-axis corresponds to the input homophily ${h}^{\mathrm{(gen)}}$ and the y-axis denotes the computed homophily $h$ based on Eq. \ref{eq:prob-h-adjusted}. Color indicates group proportions, and marker shapes indicate different minority activity rates.
            }
		\label{fig:baseline-synthetic}
	\end{center}
\end{figure}

Using the same network model for the two groups described above, we generate networks with varying input homophily values, group proportions, and activity rate disparities. In Figure \ref{fig:baseline-synthetic}, the x-axis corresponds to the input homophily ${h}^{\mathrm{(gen)}}$, the y-axis denotes the computed homophily $h$, and different colors represent different group proportions. Even if the inference method is \textit{agnostic} about the underlying generative model of the network or group's activities, there is an excellent agreement between the computed and input homophily across different values and for scenarios with different group proportions and activity differences.

\subsubsection*{Triadic Closure and Preferential Attachment}
Homophily is not the only mechanism controlling linking preferences in social networks.
Two other important processes in the study of social networks are preferential attachment and triadic closure.
Preferential attachment is the tendency of individuals to connect to those who already have a high number of connections. This process, also known as rich gets richer, helps to explain why some nodes become highly popular while others remain in the periphery, resulting in skewed degree distributions
\cite{barabasi_emergence_1999}.
Triadic closure refers to the tendency of individuals to form connections with people with whom they share mutual friends. These dynamics help to explain the formation of cliques and high link density in some areas of the network.

The interplay of homophily with preferential attachment \cite{karimiHomophilyInfluencesRanking2018} and triadic closure has a significant impact on the structure of social networks. This interplay can lead to a different number of within-group connections than would be predicted based solely on homophilic preferences \cite{asikainenCumulativeEffectsTriadic, abebe2022effect}. The interaction between these mechanisms hence complicates the accurate inference of homophily, as it further obscures the relationship between choice homophily and observed homophily.

Although these two mechanisms are not directly incorporated into our inferential model, we aim to evaluate the robustness of our model by analyzing networks exhibiting them. To do so, we work with a generalized version of the growth model, displaying both preferential attachment and triadic closure. \cite{bianconiTriadicClosureBasic2014}
The generative algorithm is outlined as follows:

\begin{enumerate}
    \item The network begins with two connected nodes of random types.
    
    \item A new node $i$ joins the system. With probability $f_a$, the node is of type $a$ and otherwise of type $b$.
    The node will establish $k^r_{init}$ new links with older nodes, where $r$ can be either $a$ or $b$, recognizing that groups can exhibit varying activity rates.
    
    \item \textit{Preferential attachment:} For its first link, the new node randomly chooses one older node to connect to. An older node $j$ of the same type will be chosen with a probability proportional to $k_j {h}^{\mathrm{(gen)}}$ and a node from a different type with a probability proportional to $k_j (1-{h}^{\mathrm{(gen)}})$. Here, $k_j$ is the current degree of the older node $j$.

    \item \textit{Triadic closure}: For the remaining $k^r_{init}-1$ links, with probability $t$, node $i$ will close a triangle by connecting to one of its neighbors' neighbors uniformly at random. With probability $1-t$, node $i$ will instead form a link following the mechanism of biased preferential attachment of step 3.

\end{enumerate}

\begin{figure}[!ht]
	\begin{center}
		\includegraphics[width=0.8\linewidth]{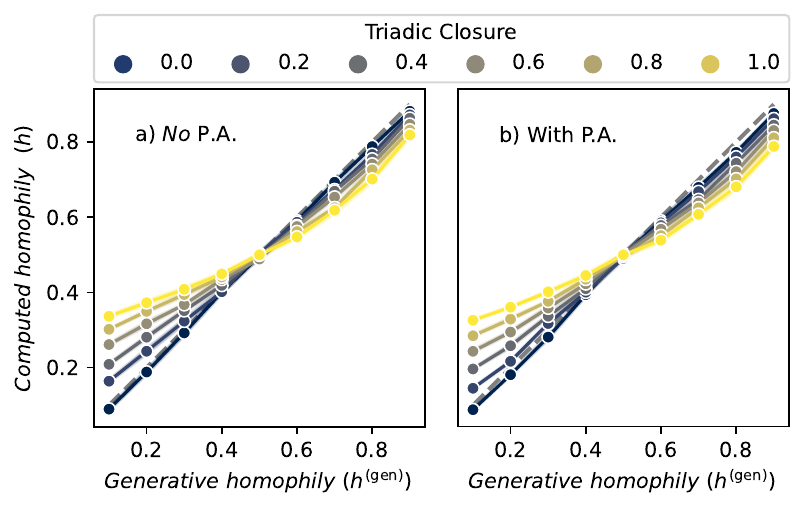}
		\caption{\textbf{Evaluating the performance of our inference method using the network model with homophily and triadic closure, a) without preferential attachment (P.A.) and b) with the preferential attachment.}
         The x-axis corresponds to the input homophily ${h}^{\mathrm{(gen)}}$, and the y-axis denotes the computed homophily $h$. Different colors represent different intensity levels for the triadic closure.
         }
		\label{fig:synthetic-no-PA}
	\end{center}
\end{figure}

The probability $t$ represents the intensity of triadic closure in the network.
We use these synthetic networks as the input of our inference model, obtaining the inferred homophily and comparing them to the input homophily values.
Figure \ref{fig:synthetic-no-PA} panels a and b respectively illustrate the results of the cases without and with preferential attachment.
Different colors represent different probabilities of triadic closure. 
The scenarios with $t=0$ correspond to the case with no triadic closure.
The results indicate that triadic closure has a significant impact on the inferred homophily, particularly in scenarios with higher input heterophily ( ${h}^{\mathrm{(gen)}} \ll 0.5 $ ). However, the deviation remains relatively small, never exceeding 0.3, even in the most extreme cases. Moreover, high triadic closure intensities do not result in significant deviations if the network is homophilic ( ${h}^{\mathrm{(gen)}} >  0.5 $ ), which is the most common case in real social networks.
Surprisingly, despite not being considered in our mathematical framework, preferential attachment does not impact the accuracy of our inference at all.
We hypothesize that by controlling for the relative activity of the groups (the number of available stubs $K_r$),
we effectively control for any degree-related connection preference. 

\subsubsection*{Asymmetric Homophily Preferences}\label{sec:directed_general}

In the earlier sections, we employed a fixed homophily preference parameter $h$ to model the preferences of the two interacting groups. However, this may not fully capture the complexities of real-world networks, as the preference for connecting with similar individuals could be asymmetric, and each group may have differing levels of preference. To account for this possibility, we introduce a generalized framework that allows different homophily preferences for each group of nodes in the network. Specifically, we consider the case of two sets of nodes, denoted by $a$ and $b$, each with their respective homophily preferences, denoted by $h_a$ and $h_b$. 

We apply this framework to directed networks, where the number of links across groups $e_{ab}$ and $e_{ba}$ can be directly observed. The derivation of the probability distribution for the directed ensemble is a straightforward generalization of the symmetric version and can be found in the Supplementary Information Section \emph{\ref{SI:microcanonical_directed_analytics}}.

To assess the validity of this generalization, we use synthetic networks with known homophily values again. These synthetic networks are created using a method similar to the one used so far while introducing two distinct homophily values, $h_a^\mathrm{(gen)}$ and $h_b^\mathrm{(gen)}$, attributed to individuals from each group. These values control the connection probabilities of the new nodes joining the system.
Additionally, the links are interpreted as directed, always originating from the new node to the older one.
These adjustments provide us with a synthetic directed network with asymmetric homophily values. Figure \ref{fig:SI:directed_inference} in SI shows the results of such an experiment using homophily values $h_a^\mathrm{(gen)} = 0.4$, $h_b^\mathrm{(gen)} = 0.6$ with minority fraction $f_a=0.2$ for a network of $1000$ nodes. As we can observe, the maximum likelihood estimation given by our framework correctly recovers the input homophily values.

\subsection*{Canonical Ensemble Equivalent}\label{sec:canonical}
Although we have derived an explicit expression for the likelihood function, we still need numerical methods to infer the value of $h$, as there is no closed-form expression for the normalization constant of Eq. \eqref{eq:prob-h-adjusted}. We can circumvent this issue and obtain a closed-form expression for $h$ by relaxing some of the assumptions of the model.
The model presented above assumes a fixed degree value for each node in the network, making it a microcanonical ensemble. If we relax these constraints, we can introduce a canonical counterpart that preserves the ensemble average of the degrees instead of their exact values \cite{peixoto2017nonparametric}.

In this canonical model, we characterize the network by its adjacency matrix $A_{ij}$, where $A_{ij}$ is the number of links between nodes $i$ and $j$ if $i \neq j$, and double that number if $i=j$. We then consider each element of the adjacency matrix as a random variable that depends on three latent variables: the activities $\theta_i$ and $\theta_j$ of nodes $i$ and $j$, which are directly related to the nodes' degrees, and the connection preference $\lambda_{\beta_i, \beta_j}$ between $i$'s group $\beta_i$ and $j$'s group $\beta_j$. In particular, we impose that the expected value of $A_{ij}$ is $\theta_i\theta_j\lambda_{\beta_i \beta_j}$.
With this constraint, $A_{ij}$ is given by a Poisson distribution with parameter $\theta_i\theta_j\lambda_{\beta_i \beta_j}$ as stated in Eq. \ref{eq:canonical:prob_link}.

\begin{equation} \label{eq:canonical:prob_link}
P(A_{ij}|\boldsymbol{\lambda},\boldsymbol{\theta}) = \frac{(\theta_i\theta_j\lambda_{\beta_i \beta_j})^{A_{ij}}e^{-\theta_i\theta_j\lambda_{\beta_i \beta_j}}}{A_{ij}!}
\end{equation}

Intuitively, we conceptualize $A_{ij}$ as the number of edges between the pair of nodes $i$ \& $j$.
Note that $A_{ij}$ can take values greater than 1, allowing for multi-edges. We can either keep the multi-edges or collapse the multigraph into a simple graph.

To determine the values of $\theta$ and $\lambda$, we calculate the ensemble average of the elements of the mixing matrix $e_{rs}$. We define $e_{rs}$ as the number of links going from block $r$ to block $s$. For $r \neq s$, this corresponds to the number of links connecting blocks $r$ and $s$, while for $r = s$, it corresponds to double the number of links within block $r$ \cite{peixoto2017nonparametric}. Thus, we have:

\begin{equation}
    \langle e_{rs} \rangle = \langle\sum_{i\in r, j \in s} A_{ij}\rangle = \sum_{i\in r, j \in s}\langle A_{ij}\rangle = \sum_{i\in r, j \in s} \lambda_{\beta_i \beta_j}\theta_i\theta_j = \lambda_{r s} \sum_{i \in r} \theta_i\sum_{j \in s}  \theta_j = \lambda_{r s} \hat{\theta}_r \hat{\theta}_s
\label{eq:canonical_mixmatrx}
\end{equation}

Where $\hat{\theta}_r$ is the sum of the $\theta$ values of all nodes belonging to block $r$. Since $\theta$ and $\lambda$ always appear together, and only their multiplication affects the probability, we cannot determine their values independently (the model is unidentifiable). However, we can make the model identifiable by imposing some reasonable constraints to obtain informative values. Choosing these constraints adequately is a critical step in obtaining a homophily measure equivalent to our microcanonical model. Since preferences in the canonical model are encoded in the $\lambda$ matrix, we need its elements to have the form:

\begin{equation}
    \lambda_{rs} = h\delta_{rs} + (1-h)(1 - \delta_{rs})
\label{eq:constraint_canonical}
\end{equation}
Where $\delta_{rs}$ is the Kronecker delta. With these constraints, we can determine the values of $h$ by plugging the values of $\lambda$ of Eq. \eqref{eq:constraint_canonical} into Eq. \eqref{eq:canonical_mixmatrx}. We then assume that the ensemble average of the mixing matrix is the empirical value that we obtained in the specific network we want to analyze: $\langle e_{rs} \rangle = \widetilde{e_{rs}}$. In a simplified case with only two blocks $a$ and $b$ this leads to the following set of equations:
\begin{equation}
\begin{aligned}
&\hat{\theta}_{a} ^ 2 h = \widetilde{e_{a a}}
\\
&\hat{\theta}_{a} \hat{\theta}_{b} (1 - h) = \widetilde{e_{a b}}
\\
&\hat{\theta}_{b} ^ 2 h = \widetilde{e_{b b}}
\end{aligned}
\label{eq:canonical:mean-degree-two-groups-3}
\end{equation}

By solving the equations for $h$, we obtain:

\begin{equation}
    h = \frac{1} { 1 + \frac{\widetilde{e_{a b}}}{\sqrt{\widetilde{e_{a a}} \widetilde{e_{b b}}}} }    
\end{equation}

We have also derived the probability distribution for $h$ given the number of inter ($\widetilde{e_{a b}}$) and intra-group links ($\widetilde{e_{a a}}$ and $\widetilde{e_{b b}}$), which enables the computation of confidence intervals. For the details, see Supplementary Information Section \emph{\ref{SI:canonical-distriubtion}}.

The equations for the directed case with asymmetric homophily values $h_a$ and $h_b$ can be obtained using a similar approach:

\begin{align}
    h_a = \frac
{ \frac{N_a}{N_b} \frac{  \widetilde{e_{aa}} }{  \widetilde{e_{ab}} } }
{ 1 + \frac{N_a}{N_b} \frac{  \widetilde{e_{aa}} }{  \widetilde{e_{ab}} } }  ,&\
 h_b = \frac
{ \frac{N_b}{N_a} \frac{  \widetilde{e_{bb}} }{  \widetilde{e_{ba}} } }
{ 1 + \frac{N_b}{N_a} \frac{  \widetilde{e_{bb}} }{  \widetilde{e_{ba}} } }
\end{align}

The full derivation for the directed networks is detailed in Supplementary Information Section \emph{\ref{SI:canonical_directed_analytics}}.

While microcanonical and canonical ensembles are not equivalent in general, their differences quickly become negligible as the network size increases. Our numerical experiments show that the canonical version offers good results in almost every situation, and its predictions are generally indistinguishable from the microcanonical model. For an evaluation of the performance of both models, see Supplementary Information Section \emph{\ref{sec:canon-vs-micro}}.
Since the canonical ensemble model is computationally cheaper, we use it instead of the microcanonical for the remaining computations presented in the paper.

\subsection*{Inferring choice homophily in empirical networks}

To illustrate the usefulness of our inference method for analyzing real-world networks, we apply it to four datasets, two containing information about friendship ties in high schools and two about co-authorship in scientific publications. In these four networks, we look at (binary) gender homophily. The four datasets are described in Supplementary Information Section \emph{\ref{SI:empirical}}.

Since the datasets on scientific collaborations span through a long time period, we build the networks by splitting the data in non-overlapping consecutive temporal windows and considering only the papers published within each period. In this way, each network is considered as one independent snapshot of a collaboration network. 
Using our inference framework, we obtain the homophily of the network in each interval. In the case of school friendships, we look at friendship ties across genders at the school level.

We evaluate the choice homophily obtained using our framework by comparing it to a traditional homophily measure. Such a measure should account for group size and activity disparities without conflating their impact on the choice homophily.
As shown in our initial comparison between outcome and choice homophily, the EI index does not account for the effect of group size disparities. As an alternative, we could use Coleman's homophily index, which was originally conceived to control for imbalanced populations. However, Coleman's index does not control for activity disparities, and it additionally suffers from other issues that we discuss in Supplementary Information Section \emph{\ref{sec:coleman-outcom-vs-choice}}. Thus, we compare our inferred homophily measure to Newman's assortativity metric \cite{newmanMixingPatternsNetworks2003}, which aims to control for group activity disparities using a configuration model-like correction to the proportion of within-group link counts. Newman's assortativity is defined as $\frac{Tr (\pmb{\epsilon}) - ||\pmb{\epsilon} ^ 2||}{ 1 - ||\pmb{\epsilon} ^ 2||}$, where the element $\epsilon_{rs}$ denotes the fraction of all edges in the network that connect nodes of type $r$ to nodes of type $s$, and $Tr(\pmb{\epsilon})$ and $ ||\pmb{\epsilon}|| $ respectively indicate the sum of the diagonal elements, and the sum of all elements of the matrix $\pmb{\epsilon}$.

\begin{figure}[!ht]
	\begin{center}
		\includegraphics[width=1\linewidth]{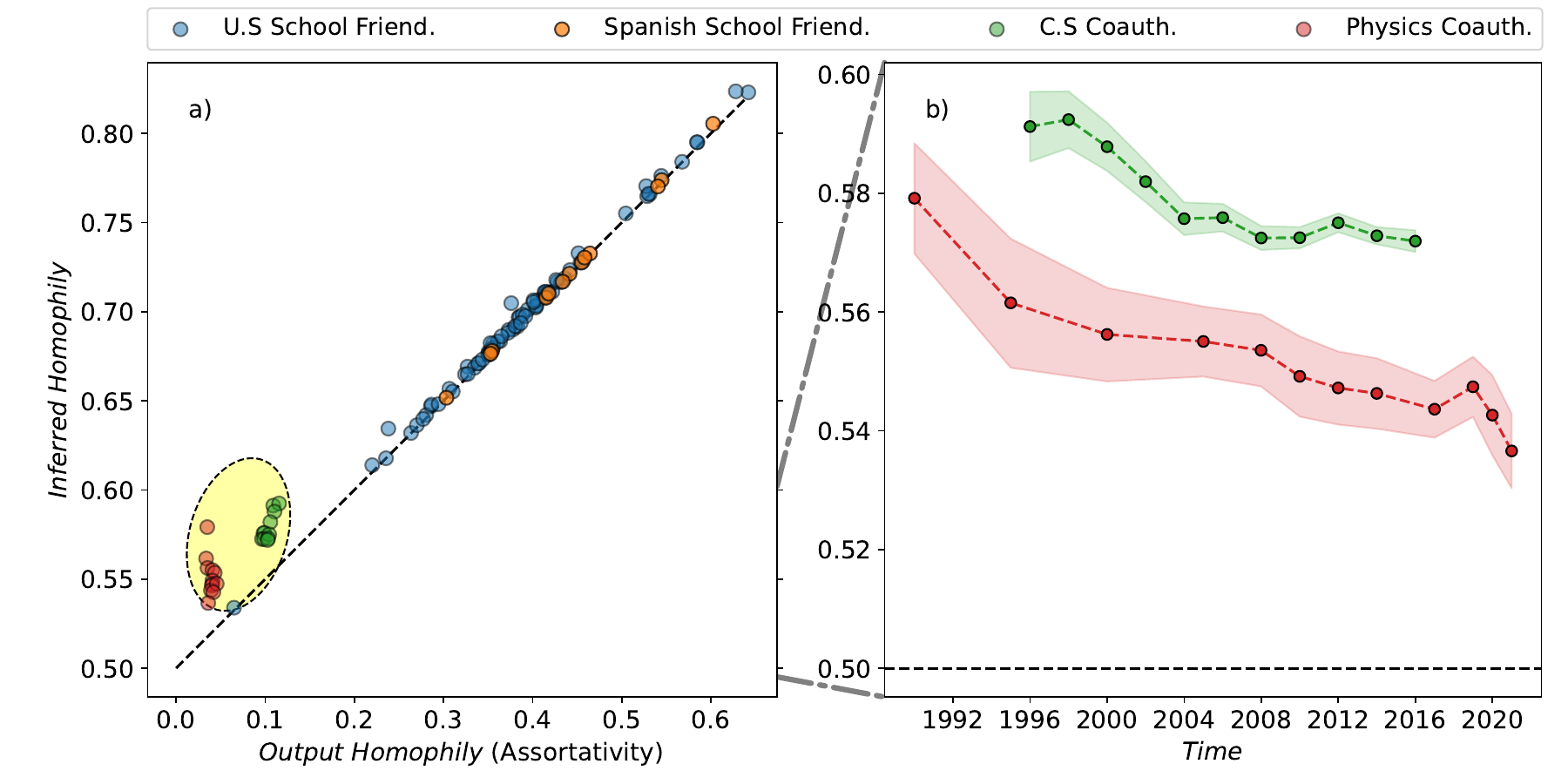}
		\caption{ \textbf{Inferred homophily for empirical networks.} a) Comparison between the outcome homophily (Newman assortativity) and the inferred choice homophily for empirical networks. The color indicates the data sources. Each point indicates a school in the friendship datasets and the time interval aggregated for the co-authorship datasets.
        b) Inferred homophily of the C.S and Physics co-authorship networks over time. The x-axis indicates time, and the y-axis denotes the inferred homophily.
        The shades indicate the $95\%$ confidence interval. The gray dash-dotted lines are guides to demonstrate the difference in scales of the two panels.
  }
		\label{fig:empirical-choice-vs-newman}
	\end{center}
\end{figure}

Figure \ref{fig:empirical-choice-vs-newman} panel a) shows the comparison between Newman assortativity and our inferred homophily in the four datasets. First, we observe two distinct behavioral patterns. In the case of the school friendship networks, there is a linear relationship between choice homophily and Newman's assortativity. However, this correlation does not hold for academic collaborations. The reason is that genders are almost perfectly balanced in schools, while the proportion of women in the scientific collaboration datasets is consistently lower than men's and fluctuates with time (see SI Figure \ref{fig:SI:empirical-assortativity-inference-proportion}). As a result, Newman's assortativity only reflects choice homophily when group proportions and activities are balanced, as seen in schools with respect to gender composition. However, when groups are imbalanced in terms of size and activity, assortativity fails to reflect changes in choice homophily, as exemplified by the co-authorship networks. Lastly, our inferred homophily values in imbalanced networks are consistently higher than Newman's, suggesting that choice homophily may go unnoticed if an inadequate metric is used. These results corroborate similar recent findings regarding a positive gender-based homophily in collaborations across a heterogeneous scholarly landscape \cite{wang2023gender,karimi2022inadequacy}.

Considering the temporal aspect of the co-authorship networks, they offer a way to systematically explore gender choice homophily across time.
The results are illustrated in Figure \ref{fig:empirical-choice-vs-newman} panel b), where we observe a small but significant homophilic preference in both Computer Science and Physics, reflecting the scientists' preference to collaborate with individuals of the same gender.
Interestingly, this bias decreases over time in both fields, gradually approaching a more neutral scenario ($h=0.5$) as time progresses. Other homophily metrics are unable to capture this behavior, as shown in Supplementary Figure \ref{fig:SI:empirical-assortativity-inference-proportion}.

\section*{Discussion}

In this study, we develop a robust methodology to infer choice homophily in social networks. Our approach utilizes statistical network ensembles to estimate and standardize the measurement of individuals' preferences for connecting with similar others, taking into consideration the structural features of networks. We devise a principled measure of connection preference based on counting the number of possible network configurations with a given number of inter-group links. We explicitly incorporate inter-group linking bias as the probability weight of each configuration. The result is a statistical model that intrinsically controls for group size imbalance and disparity of social connectivities (or activity), which are structural sources of homophilic behavior unrelated to choice or preference. Through the derivation of a canonical ensemble approximation, we not only find a closed-form expression for the homophily metric but also provide confidence intervals, thus allowing for robust error estimation.

To validate the effectiveness of our approach, we conduct experiments using synthetic networks. These networks are generated with known generative homophily, allowing us to assess the ability of our model to recover the underlying homophilic patterns. Our findings show that, while traditional homophily metrics fail to recover the input homophily, our inference method is robust and reliable. We find that our model accurately captures generative homophily even in networks generated with mechanisms that it does not incorporate explicitly. Specifically, our measure remains unaffected by the presence of preferential attachment. This means that by controlling for group-wise activity disparities, we effectively control for any degree-related connection preference. As a caveat, our measure is influenced to some extent by triadic closure, indicating that the impact of this mechanism on homophily estimation requires further investigation. However, the measurement error is relatively small in homophilic systems, which are prevalent in real-world social contexts. 

Applying the inference method to real scientific collaboration networks and high school friendships, we find that in systems with balanced gender populations, such as schools, our proposed method aligns closely with traditional metrics. However, when there are imbalances in group size and activity, as observed in scientific collaboration networks, our homophily measure provides distinct insights that deviate from standard metrics, emphasizing the importance of considering structural factors. This highlights the limitation of traditional homophily metrics in capturing the underlying preference-based associations and the usefulness of the proposed methodology to measure homophily in real-world networks. 

Overall, our work offers a comprehensive and principled framework for quantifying choice homophily in social networks, addressing limitations in existing methodologies. By controlling for disparities in group size and social resources and providing error estimation, we provide an efficient, precise, and robust method of measuring homophily.
Our method lays a solid foundation for standardizing the measurement of homophily and opens up new avenues to revealing homophilic patterns that may have been overlooked or reevaluating those that may have been inaccurately estimated.

\section*{Acknowledgment}

S.S. and S.MG. were supported by the Austrian research agency (FFG) under project No. 873927 ESSENCSE. F.K. was partly supported by the EU Horizon Europe project MAMMOth (Grant Agreement 101070285). We thank Network Inequality group at Complexity Science Hub Vienna for their feedback.

\section*{Code Availability Statement}
The simulation and analysis are conducted by \textit{homoph-infer} software written in Python, developed by S.S, available on
\href{https://github.com/Sepante/homoph-infer}{\textcolor{blue}{https://github.com/Sepante/homoph-infer}}
under GPLv3.

\clearpage

\bibliography{homophily_inference.bib}

\onecolumn

\clearpage
\setcounter{section}{0}
\setcounter{subsection}{0}


\renewcommand\thesection{S\arabic{section}}

\begin{center}
\textbf{\large Supplementary Information}
\end{center}
\setcounter{equation}{0}
\setcounter{figure}{0}
\setcounter{table}{0}
\setcounter{page}{1}
\makeatletter
\renewcommand{\theequation}{S\arabic{equation}}
\renewcommand{\thefigure}{S\arabic{figure}}
\renewcommand{\thetable}{S\arabic{table}}

\section{Coleman's index and group size adjustment} \label{sec:coleman-outcom-vs-choice}

In the main document, we illustrated the relationship between the choice and outcome homophily using the EI measure to quantify outcome homophily. In this section, we show the same discrepancy using Coleman's index \cite{colemanRelationalAnalysisStudy1958}.
To calculate this index, we use $K_r$ as the total number of stubs of agents of type $r$.
Following the convention, we use ${e_{rr}}$ as the number of stubs between agents of type $r$ (double the number of links). Hence, $H_r = \frac{e_{rr}}{K_{r}}$ represents the proportion of stubs of type $r$ that end up connecting to another node of type $r$. Coleman's index $C_r = \frac{H_r - f_r}{ 1 - f_r }$ is defined using the normalization of this parameter based on $f_l$, the fraction of the population of the nodes of type $r$.

In the case of a network with two types of nodes, Coleman's index provides us with two different values of homophily, one for the minority $a$ and another for the majority group $b$.

\begin{figure*}[!ht]
	\begin{center}
		\includegraphics[width=0.5\linewidth]{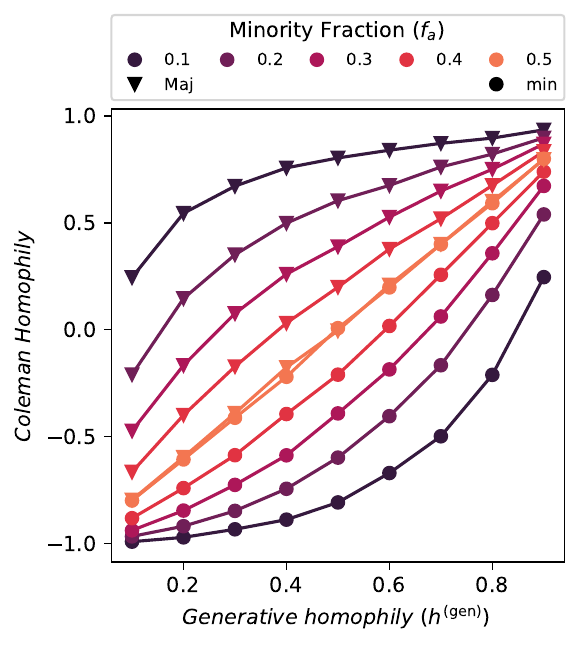}
		\caption{
Relationship between the network generative homophily and Coleman's index under different group sizes. Color denotes the fraction of the minority group.
The x-axis indicates the generative homophily value $h$, and the y-axis represents the outcome of Coleman's index.
Markers denote the group experiencing specific values of homophily, where triangles and circles, respectively, denote the homophily experienced by the majority and the minority nodes.
}
		\label{fig:Coleman-fraction}
	\end{center}
\end{figure*}

As we did for our comparison between EI and choice homophily, we generate networks with input homophily ranging from ${h}^{\mathrm{(gen)}}=0.1$ to ${h}^{\mathrm{(gen)}}=0.9$. We also vary the group sizes from the minority fraction $f_a = 0.1$ to $f_a = 0.4$. The results are illustrated in Figure \ref{fig:Coleman-fraction}. As we can see, Coleman's index for the minority and majority report lower and higher homophily, respectively.

While Coleman's index is expected to account for group size disparity, there are several reasons why it doesn't recover the input values of the synthetic model.
First, the underlying generative model behind Coleman's index is as follows: with a probability $h$, a node will connect to another node of the same type, and with $1-h$, it forms a link randomly. This choice enforces a connection with the members of the same group with probability $h$ without any consideration for the sizes of the respective groups. Only in the random case is the linking proportional to group sizes.
Second, while capturing homophily from the perspective of a particular group, it considers the link formation process one-sided and without the involvement of the link recipient.
Third, it accounts for group disparity solely based on the group size proportions.

However, the visibility of a group is not only dependent on its size. A small group of active nodes can have high visibility by forming many connections. Members of such a group could have a disproportional number of links, increasing the number of links with other groups and appearing as heterophilic even in a homophily-neutral scenario.
See section \hyperref[sec:outcom-vs-choice]{\emph{Illustrating choice vs outcome homophily}} in the main manuscript for a detailed discussion.

\begin{figure*}[!ht]
	\begin{center}
    \includegraphics[width=\linewidth]{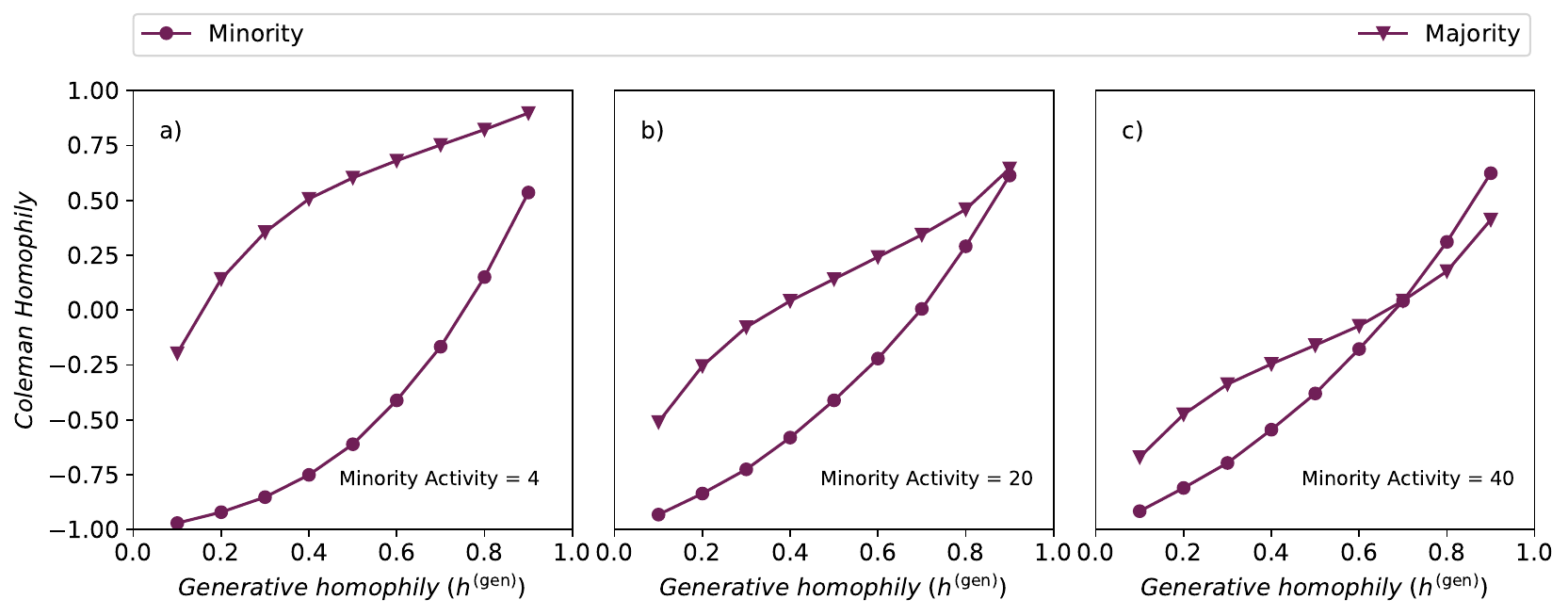}
    \caption{
            Relationship between the network generative homophily and Coleman's for different minority activity rates when majority activity is set to $4$, and the minority fraction is $f_a = 0.2$. 
            The x-axis indicates the generative homophily value $h$, and the y-axis represents the outcome of Coleman's index.
            Markers denote the group experiencing specific values of homophily.
            }
		\label{fig:SI:coleman-deg}
	\end{center}
\end{figure*}

In the following, we analyze the effect of this activity disparity between the groups on Coleman's index. 
By changing the activity rate (degree) of the minority group, in Figure \ref{fig:SI:coleman-deg} we observe that Coleman's index is also highly sensitive to this disparity. With higher values of initial connections for the minority group, the gap between Coleman's indices for the two groups gets smaller until both lines cross. It is worth noting that Coleman's index overestimates homophily for the majority and underestimates it for the minority when both groups have the same activity rates. In the case with very different activity rates, Coleman's index underestimates preferences for both groups simultaneously, categorizing them as more heterophilic than they really are.

\section{Sensitivity of the microcanonical and canonical ensembles to network density} \label{sec:canon-vs-micro}

\begin{figure*}[!ht]
	\begin{center}
    \includegraphics[width=0.6\linewidth]{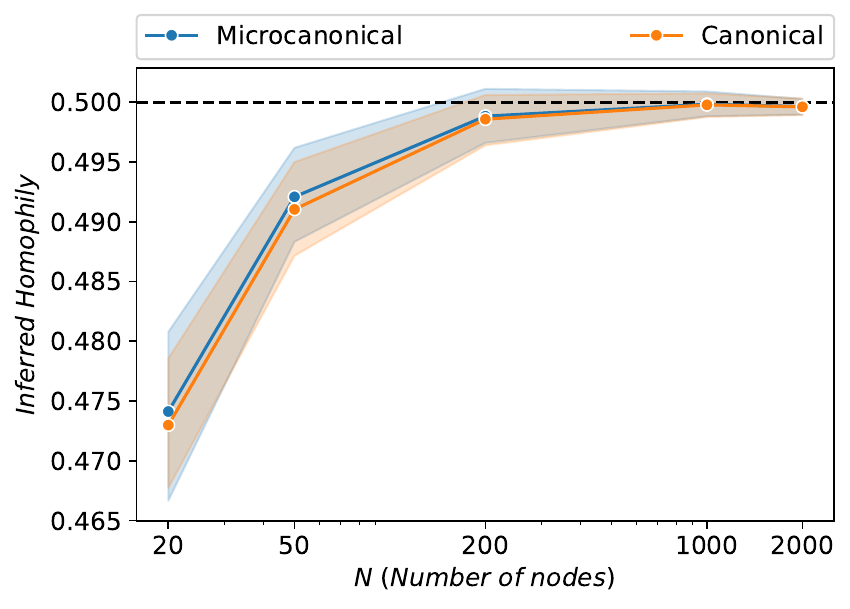}
    \caption{
            Relationship between the inferred homophily value accuracy and network size for balanced groups. The dashed line indicates the generative homophily values $h=0.5$.
            The shade indicates $95 \%$ confidence interval based on $200$ simulations.
            }
		\label{fig:SI:canon-vs-micro}
	\end{center}
\end{figure*}

In this section, we test the performance of both the canonical and the microcanonical models for networks of different sizes.
To reduce the biases caused by the generative model, here we use a generative model closer to the inference method, where a couple of links randomly rewire based on a Metropolis–Hastings algorithm, allowing two links to rewire with probability $h$ for homophilic and $1-h$ for heterophilic links.
In Figure \ref{fig:SI:canon-vs-micro}, we plot the inferred value of homophily with networks with generative $h=0.5$. The group sizes are equal, and the activity for both groups is set to $4$. Since we fix the average degree, the density of the network decreases with network size.
As we can observe, although there is a drop in the performance of the models when using smaller (denser) networks, both models recover the generative homophily value with high precision, even in these cases (notice that the inferred homophily is at most at a distance of $0.025$ from the true value).

\section{Empirical analysis using output homophily measures}\label{SI:empirical}
\subsection*{Data Description}
Here, we describe the four datasets used in the analyses:

\begin{enumerate}
    \item C.S Collaborations: Scientific collaborations among authors in the DBLP (Digital Bibliographic Library Browser) computer science bibliographic database \cite{ley2002dblp}. Each co-authored paper creates an undirected edge between the authors involved. The gender-annotated data is derived from a previous work \cite{jadidi2018gender}.
    \item Physics Collaborations: Collaborations among authors of papers published in journals of the APS (American Physical Society) between 1940 and 2010s. As in C.S, two researchers are connected if they co-author a paper during this time period. We use the gender labels inferred in previous works \cite{karimi2016inferring,kong2022influence}.
    \item U.S School Friendships (Add Health): The data is a network of high school friendships collected in 1994. It includes information from over 90,000 students in 84 communities \cite{ADD-HEALTH-moody2001peer}. Friendship ties are originally one-directional, as survey respondents were asked to nominate schoolmates they were friends with. We only consider the bidirectional links to build an undirected reciprocal friendship network.
    \item Spanish School Friendships: The data consists of social networks from 13 schools, involving more than 3,000 students and approximately 60,000 positive and negative relationships \cite{spanish-schools-ruiz2023triadic}. We only consider the positive bidirectional links to build an undirected friendship network.
\end{enumerate}

\subsection*{Collaboration networks temporal analysis}
\begin{figure*}[!ht]
	\begin{center}
		\includegraphics[width=0.5\linewidth]{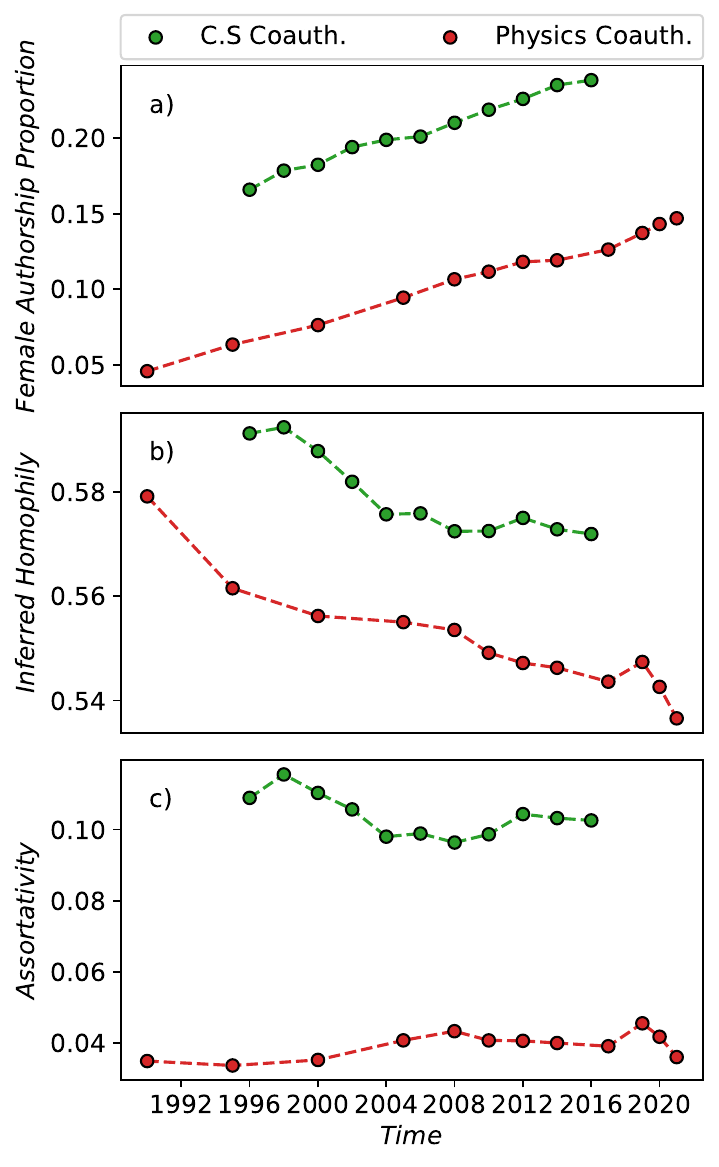}
		\caption{C.S and Physics collaboration datasets over time. a) Female authorship fraction, b) Inferred Homophily, c) Newman's assortativity.
                }
        		\label{fig:SI:empirical-assortativity-inference-proportion}
	\end{center}
\end{figure*}

In this section, we analyze the C.S and Physics collaboration datasets across time. In Figure \ref{fig:SI:empirical-assortativity-inference-proportion} a), we illustrate how the fraction of female authorships (author per paper) is linearly increasing over time.
In panel b), we observe that the inferred homophily is decreasing over time, while Newman's assortativity in panel c) is incapable of detecting this trend.

\section{Microcanonical Inference mathematical formulations}\label{SI:mathematics}

Here, we derive the probability distributions for our homophily inference framework.
\subsection*{Two groups}
\subsubsection*{Undirected Ensembles}
As the first step, we derive the probability distribution for the case in which connections are not biased (Eq. \ref{eq:neutral-prob-dist}).

\begin{equation}\label{eq:SI:2}
P({e_{ab}}|K_a, K_b) = \frac{ \Omega({e_{ab}}|K_a, K_b) }{ \Omega(K_a, K_b) } = \frac{ {e_{ab}}! \binom{K_a}{e_{ab}} \binom{K_b}{e_{ab}} (K_a - {e_{ab}}-1)!! (K_b-{e_{ab}}-1)!!  }{ (K_a+K_b-1) !! }
\end{equation}

The numerator counts the number of ways we can pair up the stubs while having exactly ${e_{ab}}$ inter-group links bridging the two groups. For this purpose, we first choose ${e_{ab}}$ stubs from each side to form the inter-group links, which for each group $r$ can be done in $\binom{K_r}{e_{ab}}$ ways. We will have ${e_{ab}}!$ ways of matching the chosen stubs. The rest of the stubs from each group $r$, ($K_r- {e_{ab}}$) have to be independently put into pairs to form the intra-group links, which can be done in $ (K_r - {e_{ab}} - 1 )!! $ different ways. The sign ($!!$) is the double factorial, and $n!!$ indicates the product of all the natural numbers up to $n$ that have the same parity.
The denominator counts the number of ways we can pair all the links $E = K_a + K_b$.

The incorporation of $h$ consists of weighting the existence of each intra and inter-group link by $h$ and $1-h$, respectively. Hence, for a configuration with ${e_{ab}}$ inter-group links, the numerator will be adjusted by a factor of $(1-h) ^ {e_{ab}} h ^ {(E - {e_{ab}})}$. The denominator would be the sum of the numerator for all possible values of ${e_{ab}}$. The final distribution would result in the following:

\begin{equation}\label{eq:SI:prob-h-adjusted}
P({e_{ab}}|K_a, K_b; h) = \frac{ \Omega({e_{ab}}|K_a, K_b; h) }{ \Omega(K_a, K_b; h) } = \frac{ {e_{ab}}! \binom{K_a}{e_{ab}} \binom{K_b}{e_{ab}} (K_a - {e_{ab}}-1)!! (K_b-{e_{ab}}-1)!! (1-h) ^ {e_{ab}} h ^ {(E - {e_{ab}})} }
{ \sum_{e_{ab}} \Omega({e_{ab}}|K_a, K_b; h) \delta_{mod(E - {e_{ab}},2), 0} }
\end{equation}

We use the Kronecker delta $\delta_{mod(E - {e_{ab}},2), 0}$ because only inter-group link counts with the same parity as the total number of edges are acceptable. Otherwise, the number of intra-group stubs would be odd, and pairing would not be possible (there would always be a loose stub).

\subsubsection*{Directed Microcanonical Ensembles}\label{SI:microcanonical_directed_analytics}
For the directed case, we define $e_{rs}$ as the number of links going from group $r$ to $s$. Notice that, unlike the undirected case, we do not double-count the ingroup links.
Hence, we have,
$K_a =  e_{aa} + e_{ab}$,
$K_b =  e_{bb} + e_{ab}$.
We define the total inter-group links $m = e_{ab} + e_{ba}$. From the total number of links for each group $K_a, K_b$, we choose $m$ total stubs from each group.
Next, we arrange the pairings of inter-group links ($m!$) and intra-group links $\binom{K_a}{e_{ab}}, \binom{K_b}{e_{ab}}$. We then count the number of arrangements based on the directions. For this, we choose $e_{ab}$ out of $m$ to be the ones directing from $a$ to $b$. Every intra-group link, on the other hand, has two possible directions, which are accounted for by powers of two. Finally, since the network is directed, we may have two distinct probabilities $h_a$ and $h_b$ associated with the in-group preferences of groups $a$ and $b$, respectively.
The normalization factor is similar to the case of Eq. \ref{eq:SI:prob-h-adjusted}, although we sum over both $e_{ab}$ and $e_{ba}$ and check for the parity within each group independently. The resulting distribution is the following:

\begin{equation}\label{eq:SI:directed-prob-h-adjusted}
\begin{split}
P(e_{ab}, e_{ba}|K_a, K_b; h_a, h_b) &=
\frac{ \Omega(e_{ab}, e_{ba}|K_a, K_b; h_a, h_b) }
{ \Omega(K_a, K_b; h_a, h_b) }\\
&=
\frac{ m! \binom{K_a}{m} \binom{K_b}{m} (K_a - m - 1 )!! (K_b-m - 1)!!  }
{ \sum_{e_{ab}, e_{ba}} \Omega(e_{ab}, e_{ba}|K_a, K_b, h_a, h_b) \delta_{mod(K_a - m, 2), 0} \delta_{mod(K_b - m, 2), 0} }\\
&\times \binom{e_{ab}}{e_{ab}}
 h_a ^{e_{aa}} (1-h_a) ^ {e_{ab}} h_b ^{e_{bb}} (1-h_b) ^ {e_{ba}} 2^{(e_{aa} + e_{bb})}
\end{split}
\end{equation}

It is worth noting that in Eq. \ref{eq:SI:directed-prob-h-adjusted}, we have maintained degree sums $K_a$ and $K_b$ as constraints.
However, it is also possible to explore an alternative formulation. This involves considering the outgoing links, which are represented as $K^{o}_a$ and $K^{o}_b$, as the constraints. This adjustment allows for variations in $K_a$ and $K_b$. Since each group allocates its outgoing links independently, the distribution can be expressed as two separate and independent distributions, as shown in Eq. \ref{eq:directed_growth_splitting}. In this formulation, each individual distribution follows a binomial distribution pattern, as represented in Eq. \ref{eq:directed_growth}.

\begin{equation}\label{eq:directed_growth_splitting}
\begin{split}
P(e_{ab}, e_{ba}|K^{(o)}_a, K^{(o)}_b; h_a, h_b) = P(e_{ab}|K^{(o)}_{a}; h_{ab}) P(e_{ba}|K^{(o)}_{b}; h_{ba})
\end{split}
\end{equation}

\begin{equation}\label{eq:directed_growth}
\begin{split}
P(e_{ab}|K^{(o)}_{ab}; h_{ab}) =
\binom{K^{{(o)}}_a}{e_{ab}} \rho_{a}^{e_{aa}} (1 - \rho_{a})^{e_{ab}} 
\end{split}
\end{equation}

Where $\rho_{a}$, $1-\rho_{a}$ denote the independent probabilities of each stub originating from group $a$ to land on its own group or on group $b$. This probability is proportional to the population sizes $n_a$, $n_b$ and the preference of group $a$ towards its own group ($h_a$) and the other group ($1-h_a$):

\begin{equation}
\begin{split}
\rho_{a} = \frac{n_a (h_{a})}{n_a h_{a} + n_b (1 - h_{a})}
\end{split}
\end{equation}

The choice between these two approaches depends on the context. For instance, in systems like citation or follower-followee networks, where the recipient of the link does not need to invest resources to maintain it, an out-degree constraint might be more appropriate. On the other hand, in situations like friendship networks, where both the initiator and the recipient allocate resources, a constraint on the total number of links might be a better fit.

One can also devise a third approach, where both the number of total links and outgoing links (hence, also ingoing links) are constant. This formulation will result in only one degree of freedom. We use $e_{ab}$ as the free parameter with the following set of equations:
\begin{equation}
\begin{split}
K_a^{(o)} = e_{aa} + e_{ab}
\\
K_a^{(i)} = e_{aa} + e_{ba}
\\
K_b^{(o)} = e_{ba} + e_{bb}
\\
K_b^{(i)} = e_{ab} + e_{bb}
\end{split}
\end{equation}

The minimum and maximum value for $e_{ab}$ would be respectively determined by $e_{ab}^{\mathrm{(min)}} = max(0, K_a^{(o)} - K_a^{(i)})$ (minimum number of outgoing links that cannot be contained in group $a$) and $e_{ab}^{\mathrm{(Max)}} = min( K_a^{(o)}, K_b^{(i)} )$ (maximum number of outgoing links from $a$ that could be contained in group $b$), where the superscript $i$ would denote the ingoing links.
To obtain the number of possible realizations given $e_{ab}$, we count the number of ways of choosing the out and ingoing stubs from both groups and then count the number of ways to couple the stubs within and across groups. After some cleaning, we will obtain the number to be
$\Omega(e_{ab}|K_a^{(o)}, K_a^{(i)}, K_b^{(o)}, K_b^{(i)}; h_{ab}) = \frac{K_a^{(o)}! K_b^{(o)}! K_a^{(i)}! K_b^{(i)}!} {e_{aa}! e_{ab}! e_{ba}! e_{bb}!}
$
where all the parameters $e$ can be written based on $e_{ab}$. For each excessive interlinks more than the minimum value ${( e_{ab} - e_{ab}^{\mathrm{(min)}})}$, we adjust the probability weight by $1-h_{a}$ and for the possible inter-links which have not been realized ${(e_{ab}^{\mathrm{(Max)}} - e_{ab})}$, we adjust the realization probability by $h_{a}$. We obtain:

\begin{equation}\label{eq:SI:directed-in-and-out}
\begin{split}
P(e_{ab}|K_a^{(o)}, K_a^{(i)}, K_b^{(o)}, K_b^{(i)}; h_{ab}) &=
\frac{ \Omega(e_{ab}|K_a^{(o)}, K_a^{(i)}, K_b^{(o)}, K_b^{(i)}; h_{ab}) \times
h_{a}^{(e_{ab}^{\mathrm{(Max)}} - e_{ab})}
(1 - h_{a})^{( e_{ab} - e_{ab}^{\mathrm{(min)}})}
}
{
\Sigma_{e_{ab}} \Omega(e_{ab}|K_a^{(o)}, K_a^{(i)}, K_b^{(o)}, K_b^{(i)}; h_{ab}) \times
h_{a}^{(e_{ab}^{\mathrm{(Max)}} - e_{ab})}
(1 - h_{a})^{( e_{ab} - e_{ab}^{\mathrm{(min)}})}
}\\
\end{split}
\end{equation}

\begin{figure*}[!ht]
	\begin{center}
		\includegraphics[width=0.6\linewidth]{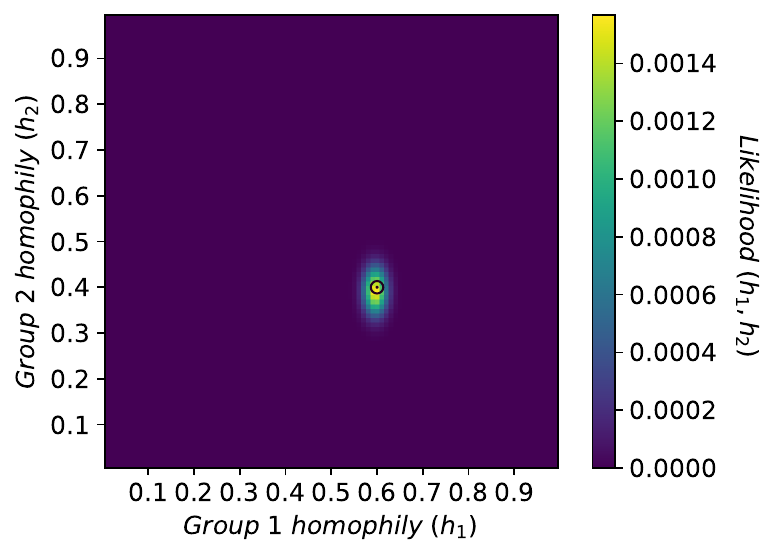}
		\caption{\textbf{Homophily inference for a synthetic directed network.}
        The colormap denotes the likelihood of the homophily inference method.
        The black ring and dot represent the generative homophily values.
            }
		\label{fig:SI:directed_inference}
	\end{center}
\end{figure*}

Since the example in section Asymmetric Homophily Preferences of the main document sets a constant number of outgoing links for each independent node, we chose to follow the second approach (Eq. \eqref{eq:directed_growth_splitting}) for this scenario as illustrated in Figure \ref{fig:SI:directed_inference}.

\subsection*{Arbitrary number of groups}\label{section:SI-more-groups}

\subsubsection*{Undirected Ensembles}
For the cases with more than two groups, we will have the sets $\{K_r\}, \{e_{rs} \},  \{ h_{rs} \} )$, respectively denoting the vector of total degree/stubs for each group $r$, the vector counting the number of links between each pair of groups $r$ and $s$, and the preference of interaction between groups $r$ and $s$.
To generalize the framework, we proceed by choosing the outgoing stubs of a group $r$ categorized by the group $s$ they will land on, using a multinomial distribution. Then, we arrange the intra and inter-group links as previously done. Finally, we adjust for homophilic preferences as in Eq. \ref{eq:SI-undirected-sym-multigroup}.

\begin{equation}\label{eq:SI-undirected-sym-multigroup}
\begin{split}
\Omega ( \{e_{rs} \} | \{K_r\}; \{ h_{rs} \} ) =
\prod^{n}_{r} \prod^{n}_{s}  \frac{K_r!}{  e_{rs}! }
(e_{rr} - 1)!!
\prod^{n}_{s | s < r} e_{rs}!
\prod^{n}_{s | s \leq r} h_{rs} ^ {e_{rs}}
\end{split}
\end{equation}

To validate the derivations, we perform a test simulating a network with three groups. The algorithm to build the network is analogous to the growth model described in the main document.
The group sizes are $N_1 = N_2 = 200$ and $N_3 = 1600$. We use the homophily values $h_{11} = h_{22} = 0.45$ and $h_{33} = 0.8$ and for simplicity set the generative across-group preference values to $h_{13} = h_{23} = 0.1$ and $h_{12} = 0.45$, resulting in the following homophily matrix:

\begin{equation}
  \pmb{h} = 
  \begin{bmatrix}
    0.45 & 0.45 & 0.1 \\
    0.45 & 0.45 & 0.1 \\
    0.1 & 0.1 & 0.8 \\
  \end{bmatrix}
\end{equation}

\begin{figure*}[!ht]
	\begin{center}
		\includegraphics[width=0.6\linewidth]{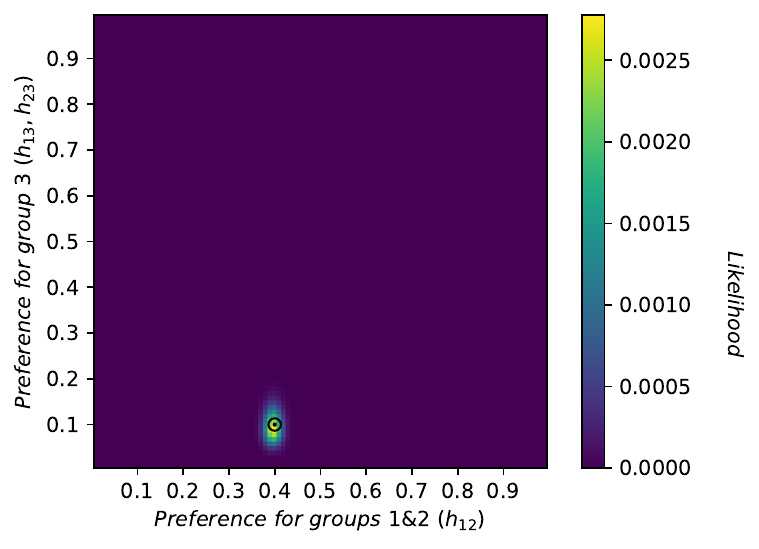}
		\caption{\textbf{Homophily inference for a synthetic multigroup network.}
  The x-axis denotes the preference between groups $1$
        The colormap denotes the likelihood of the inferred values.
        The black ring and dot represent the generative preference values.
            }
		\label{fig:SI:multigroup_inference}
	\end{center}
\end{figure*}

Hence, groups $1$ and $2$ have a lower preference for interaction with nodes of a group $3$ than interacting with nodes from each other or their own groups. The results of the inference method are illustrated in Figure \ref{fig:SI:multigroup_inference}. As can be seen, the method is capable of correctly recovering the generative preferences.

\subsubsection*{Directed Ensembles}
For the directed case, we first partition all stubs from $K_i$ into the ones going to/coming from other communities ($e_{rs}, e_{sr}$ in the denominator) and the ones remaining in the community $e_{rr}$. We then arrange the inter-group links ($e_{rs}!{h_{rs}} ^ {e_{rs}}$) and lastly we arrange the intra-group links ($( 2e_{rr} -1 )!! 2 ^ {e_{rr}} {h_{rr}} ^ {e_{rr}}$). With $2 ^ {e_{rr}}$ accounting for the two possible directions of every intra-group link.

\begin{equation}
\begin{split}
\Omega (\{K_r\}, \{e_{rs} \}; \{ h_{rr} \} ) =
\prod^{n}_{r} \frac{K_r!}
{ 2e_{rr}! \prod^{n}_{s|s \neq r} e_{rs}! e_{sr}! }
     \prod_{r,s|r \neq s}^{n} e_{rs}!  {h_{rs}} ^ {e_{rs}} 
    \prod_{r}^n ( 2e_{rr} - 1 )!! 2 ^ {e_{rr}} {h_{rr}} ^ {e_{rr}}
\end{split}
\end{equation}

\section{Canonical inference mathematical formulations}\label{SI:canonical-distriubtion}
\subsection*{Canonical Ensemble Probability Distribution}
In this section, we provide the probability distribution and, hence, the confidence intervals for the inferred homophily via the canonical ensemble.
Given our normalization assumptions in Equations \ref{eq:constraint_canonical} and \ref{eq:canonical:mean-degree-two-groups-3}, we rewrite Eq. \ref{eq:canonical:prob_link} based on $h$. To simplify the analytics, we focus on the case where nodes $i$ and $j$ are not in the same group and write the equations based on $1-h$. Hence, we have:

\begin{equation} \label{eq:SI:canonical:prob_link}
P(A_{ij}|1-h,\boldsymbol{\theta}) = \frac{(\theta_i\theta_j (1-h) )^{A_{ij}}
e^{-\theta_i\theta_j (1-h) }   }
{A_{ij}!}
\end{equation}

Summing over all the possible $i$ and $j$'s belonging respectively to blocks $a$ and $b$ we have:

\begin{equation} \label{eq:SI:canonical:prob_interlinks}
P(e_{ a b }|1-h,\boldsymbol{\theta}) = \frac{( \widehat{\theta_{a}} \widehat{\theta_{b}} (1-h) )^{e_{ a b }}
e^{ - \widehat{\theta_{a}} \widehat{\theta_{b}} (1-h) }
}
{ e_{ a b } !}
\end{equation}

We now rewrite the probability distribution to derive a likelihood for the parameter $1-h$ to produce a network with a specific value of $e_{ a b }$. After normalization, the distribution is equivalent to a Bayesian inference posterior with a uniform prior. To achieve this, we rewrite
$ \widehat{\theta_{a}} \widehat{\theta_{b}} $ as $\mu$ where $ \mu = \sqrt{e_{ a a } e_{ b  b}} + e_{ a b }$ based on Eq. \ref{eq:canonical:mean-degree-two-groups-3}.
Hence, we have:

\begin{equation} \label{eq:SI:canonical:likelihood}
P((1-h)|\boldsymbol{e},\boldsymbol{\theta}) = \frac{(  1-h )^{e_{ a b } - 1} \mu ^ {e_{ a b }}
e^{ - \mu (1-h) }
}
{ e_{ a b } !}
\end{equation}
Eq. \ref{eq:SI:canonical:likelihood} is a Gamma distribution over the random variable $1-h$ with parameters $\alpha = e_{a b}$ and $\beta = \mu$.
Hence, we can easily obtain its confidence interval using available statistical packages such as SciPy.

\begin{figure*}[!ht]
	\begin{center}
		\includegraphics[width=0.7\linewidth]{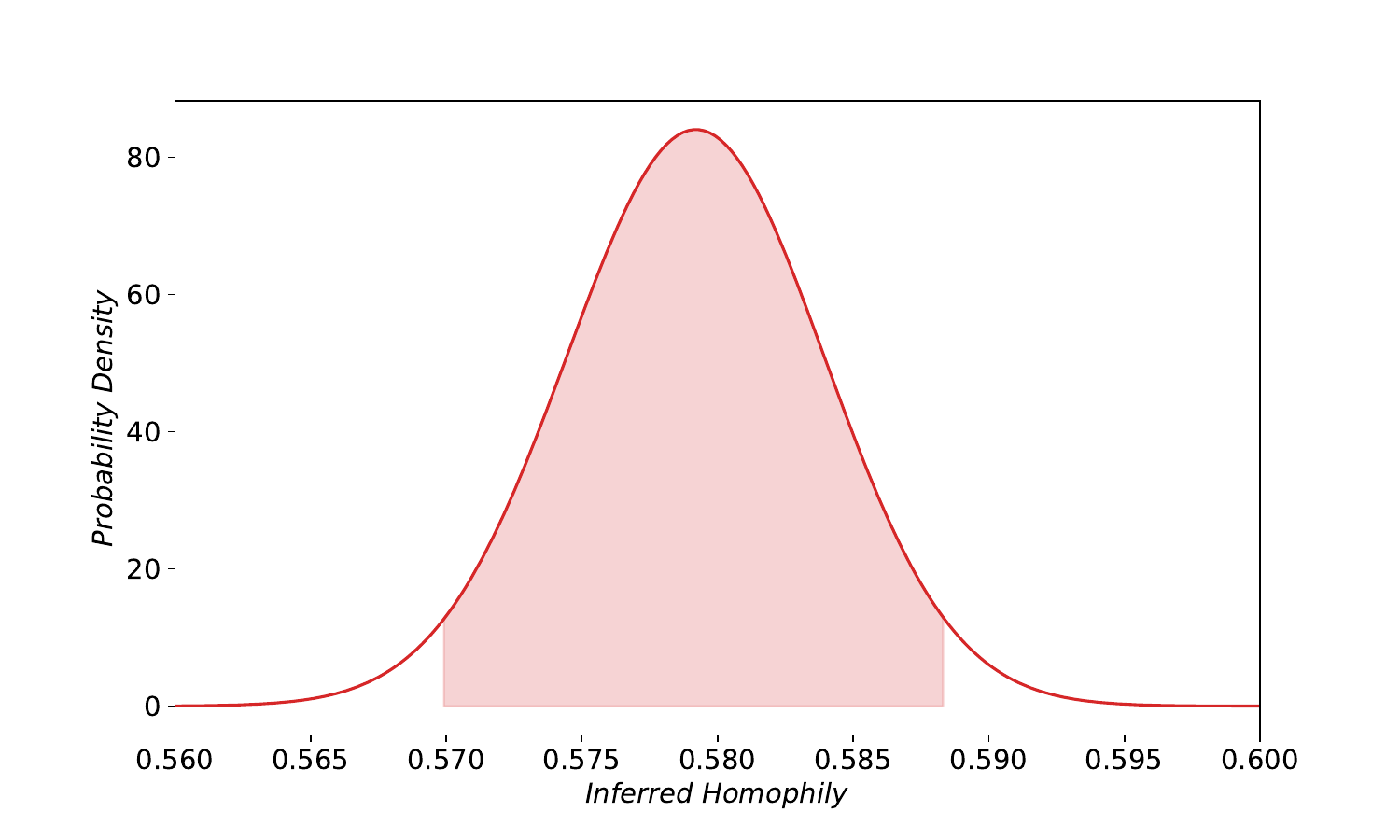}
		\caption{
Probability distribution of the inferred homophily, given by the Gamma distribution. The shade indicates the $95\%$ confidence interval.
The data represents the physics co-authorship network for the 1996-2000 interval depicted in Figure \ref{fig:empirical-choice-vs-newman}.
                }
        		\label{fig:canon-homophily-confidence}
	\end{center}
\end{figure*}

Figure \ref{fig:canon-homophily-confidence} illustrates the probability distribution of the inferred homophily, which is modeled by the Gamma distribution. The shaded region in the plot represents the 95\% confidence interval. This data pertains to the physics co-authorship network during the period of 1996-2000, also represented in Figure \ref{fig:empirical-choice-vs-newman}.

\subsubsection*{Directed Canonical Ensembles}\label{SI:canonical_directed_analytics}

In section \emph{\ref{SI:microcanonical_directed_analytics}}, we devised alternative ensembles with constraints based on degrees or out-degrees.
In the canonical ensembles we do not fix the degrees, but the expected degree or activity rates will be fixed.
Therefore, in this formulation, instead of exact degree values, we constrain the distribution based on activity rates ($\theta$).
Since the network is directed, we fix the outward activity of nodes $\theta_i^{(o)}$; that is, the average number of links source nodes will actively establish to target nodes. Since the probability of connection only depends on the activity of the node sending the link, the probability of the existence of an edge would be written as follows:

\begin{equation} \label{eq:canonical:outlink_prob_link}
P(A_{ij}|\boldsymbol{\lambda},
\boldsymbol{\theta}^{(o)})
= \frac{(\theta_i\lambda_{\beta_i \beta_j})^{A_{ij}}e^{-\theta_i^{(o)}\lambda_{\beta_i \beta_j}}}{A_{ij}!}
\end{equation}

From this distribution, we can obtain the expected values for our four observables, $\widetilde{e_{aa}}, \widetilde{e_{ab}}, \widetilde{e_{ab}}, \widetilde{e_{ba}}$. Again, $e_{rs}$ is the number of links going from block $r$ to block $s$.
As the linkage only depends on the outward propensity of the node sending the link, the expected number of links initiated by group $a$ and received by $b$ depends on the activity of group $a$ and the number of nodes in the group $b$ (each one posing as a potential candidate for a receiving link):
\begin{equation}
\begin{aligned}
    \langle e_{rs} \rangle = \langle\sum_{i\in r, j \in s} A_{ij}\rangle
    = \sum_{i\in r, j \in s}\langle A_{ij}\rangle
    = \sum_{i\in r, j \in s} \lambda_{\beta_i \beta_j}\theta_i^{(o)}
    = \sum_{j \in s} \sum_{i \in r} \lambda_{\beta_i \beta_j} \theta_i^{(o)}
    = N_s \lambda_{r s} \hat{\theta}_{r}^{(o)} 
\end{aligned}
\label{eq:canonical-directed}
\end{equation}

Using two homophily values, $h_a, h_b$ where $\lambda_{r s} = h_{r} \delta_{rs} + (1-h_{r}) (1-\delta_{rs}) $, the directed equivalent to the Eq. \ref{eq:canonical:mean-degree-two-groups-3} would be Eq. \ref{eq:canonical:mean-degree-directed}.

\begin{equation}
\begin{aligned}
&\hat{\theta}_{a}^{(o)}  h_a N_a = \widetilde{e_{a a}}
\\
&\hat{\theta}_{a}^{(o)} (1 - h_a) N_b = \widetilde{e_{a b}}
\\
&\hat{\theta}_{b}^{(o)}  h_b N_b = \widetilde{e_{b b}}
\\
&\hat{\theta}_{b}^{(o)} (1 - h_b) N_a = \widetilde{e_{b a}}
\end{aligned}
\label{eq:canonical:mean-degree-directed}
\end{equation}

Where $h_r$ and $N_r$ denote the homophily and the number of nodes, and $\hat{\theta}_{r}^{(o)}$ indicates the aggregate outward activity of group $r$.
By solving for $h_a$ and $h_b$ we have:

\begin{equation}
\begin{aligned}
h_a = \frac
{ \frac{N_a}{N_b} \frac{  \widetilde{e_{aa}} }{  \widetilde{e_{ab}} } }
{ 1 + \frac{N_a}{N_b} \frac{  \widetilde{e_{aa}} }{  \widetilde{e_{ab}} } }
\\
h_b = \frac
{ \frac{N_b}{N_a} \frac{  \widetilde{e_{bb}} }{  \widetilde{e_{ba}} } }
{ 1 + \frac{N_b}{N_a} \frac{  \widetilde{e_{bb}} }{  \widetilde{e_{ba}} } }
\end{aligned}
\end{equation}

\end{document}